\documentclass[pra,aps,twocolumn,showpacs]{revtex4}

\usepackage{graphicx}
\usepackage{amssymb}
\usepackage{graphicx}
\usepackage{bm}
\usepackage{amsmath}

\newcommand{\ra}{\rangle}
\newcommand{\la}{\langle}

\begin{document}

\title{Spatial decoherence near metallic surfaces}

\author{R.~Fermani\footnote{Electronic address: rachele.fermani@imperial.ac.uk}, S.~Scheel, and P.L.~Knight}

\affiliation{Quantum Optics and Laser Science, Blackett Laboratory,
Imperial College London, Prince Consort Road, London SW7 2BW, United Kingdom}

\date{\today}

\begin{abstract}
We present a first-principles derivation of spatial
atomic-sublevel decoherence near
dielectric and metallic surfaces. The theory is based on the
electromagnetic-field quantization in absorbing dielectric media. We
derive an expression for the time-variation of the off-diagonal matrix
element of the atomic density matrix for arbitrarily shaped
substrates. For planar multilayered substrates we find that for small
lateral separations of the atom's possible positions the spatial
coherence decreases quadratically with the separation and inversely to
the squared atom-surface distance.
\end{abstract}


\maketitle
%%%%%%%%%%%%%%%%%%%%%%%%%%%%%%%%%%%%%%%%%%%%%%%%%%%%%%%%%%%%%%%%%%%%%%
\section{Introduction}

New physical models for quantum information processing and quantum
computation have been inspired recently by the experimental
achievements in trapping and controlling ultracold neutral atoms
\cite{GREINER,NATURE_PHASETrans,SCHIEDMAYER,JAKSCH1999,ZANARDI}.
The first experimental step to achieve a physical realization of a
quantum computer with neutral atoms is to confine them on a
definite region in the space. The creation of microscopic guides
and traps for neutral atoms moving close to surfaces is possible
using nanofabricated structures that either carry currents or are
based on permanent magnetized films. The idea at the base of atom
chips has been put forward by Frisch and Segr\'{e} \cite{SEGRE}
who realized that, when a homogeneous magnetic field ('bias
field') is superimposed with the field created by a current
flowing through a wire, the magnetic field vanishes on a line
parallel to the current which can trap atoms in low-field seeking
magnetic hyperfine sublevels.

One of the main requirements for a qubit is to be well isolated
from a noisy environment to avoid decoherence, namely the
destruction of quantum superpositions due to the coupling of the
atom cloud to the noisy chip environment. Although neutral atoms
are considered good candidates as quantum systems since they have
a small coupling to the environment, they still suffer from loss and
decoherence. When atoms are trapped in atom chips, they are held
close to the material surfaces.  The small separation between the
cold atom cloud and the macroscopic environment (usually at room
temperature) raises the question of how strong the energy exchange
will be, and which limit of atom confinement and height above the
surface can ultimately be reached. Thermal fluctuations induce noise
currents \cite{JOHNSONOISE} in the materials the trap is made of, and
fluctuations of the electromagnetic field are produced in the
conducting body. Such fluctuating fields can be strong enough for
an atom close to the surface to drive rf magnetic dipole
transitions that flip its spin causing either its loss or
decoherence of its quantum state.

In
\cite{Varpula,HENKEL/99,FOLMAN&Al,HENKEL/03,SPIN-FLIP,Henkel05,Scheel05},
atom loss due to thermally driven spin flips has been widely
investigated and several experiments have confirmed the theoretical
findings \cite{EXPATOMCHIP,Vuletic,Cornell}.
In this article we examine the influence of
thermally-induced spin flips on the coherence properties of atomic
spatial superposition states.  Such coherent superpositions can be
thought of being created by tunneling through a shallow potential
barrier in either a double-well potential or, more generally, an
optical lattice structure \cite{MANDEL_nature}. The study
of the latter has been received much attention over the recent years
for its potential application in quantum information processing
(see, e.g. \cite{PACHOS03,JAKarXiv}). The derivation is carried
out within the framework of the quantum electrodynamic theory for
electromagnetic fields in dielectric media
\cite{VOGEL,AGARWAL,GRUNER,DUNG/98,SCHEEL/98,SCHEEL/99,DUNG/00,PERINA}
which yields a first-principle description of the decoherence
properties of spatial atomic superposition states.

This work is organized as follows: Sec. II introduces the basic
notions of a quantized electromagnetic field in a dielectric
medium. In Sec. III the density matrix of the atom is obtained in
the presence of a fluctuating magnetic field and an expression for
the spatial coherence is derived. We focus on a particular
substrate geometry, a planarly multilayered structured, in
Sec.~\ref{sec:planar}, for which the dyadic Green function is
explicitly known.

%%%%%%%%%%%%%%%%%%%%%%%%%%%%%%%%%%%%%%%%%%%%%%%%%%%%%%%%%%%%%%%%%%%%%%
\section{Basic equations}
\label{sec:QED}

It is well known that the quantum statistical properties of
electromagnetic fields and their interactions with atomic systems
can be strongly influenced by the presence of dielectric bodies.
In the present context it is useful to formulate quantum electrodynamics
(QED) on a dielectric-matter background
\cite{VOGEL,AGARWAL,GRUNER,DUNG/98,SCHEEL/98,SCHEEL/99,DUNG/00,PERINA}.
The interaction between atomic systems and the electromagnetic
field is typically treated in terms of the polarization and
magnetization associated with the atomic charges. Let us restrict
our attention to an isotropic but arbitrarily inhomogeneous medium
whose polarization responds linearly and locally to the electric field.
Causality and the dissipation-fluctuation theorem
\cite{FLUCT/DISSIP} then require that
\begin{equation}
\label{Polarization}
\mathbf{P}(\mathbf{r},t)= \varepsilon_0 \int\limits_0^{\infty} d\tau\,
\chi(\mathbf{r},t) \mathbf{E}(\mathbf{r}, t-\tau)
+\mathbf{P}_N(\mathbf{r},t) \,,
\end{equation}
where $\chi(\mathbf{r},t)$ is the dielectric susceptibility (in
the time domain) and $\mathbf{P}_N (\mathbf{r},t)$ is the noise
polarization associated with dissipative processes in the dielectric
medium.

Using Maxwell's equations in Fourier space, we find that
$\mathbf{E}(\mathbf{r},\omega)$ obeys the Helmholtz equation
\begin{equation}
\label{HelmholtzEq}
\bm{\nabla} \times \bm{\nabla} \times \mathbf{E}(\mathbf{r},\omega)
-\frac{\omega^2}{c^2}\varepsilon (\mathbf{r},\omega)
\mathbf{E}(\mathbf{r},\omega)=
\omega^2 \mu_0 \mathbf{P}_N(\mathbf{r},\omega),
\end{equation}
where the complex permittivity, $\varepsilon (\mathbf{r},
\omega)\!=\!\varepsilon_{R} (\mathbf{r}, \omega)+i\,
\varepsilon_{I} (\mathbf{r}, \omega)$, is defined by
\begin{equation}
\varepsilon (\mathbf{r}, \omega) =
1+\int\limits_0^{\infty} d \tau e^{i\omega t} \chi (\mathbf{r},\tau).
\end{equation}
The solution to Eq.~(\ref{HelmholtzEq}) can then be written as
\begin{equation}
\label{ElectricGreen}
\mathbf{E}(\mathbf{r},\omega) =
\omega^2 \mu_0 \int d^3 \mathbf{r}'\bm{G}(\mathbf{r},\mathbf{r}',\omega)
\mathbf{P}_N(\mathbf{r}',\omega),
\end{equation}
where the Green tensor $\bm{G}(\mathbf{r},\mathbf{r}',\omega)$
is a second rank tensor that has to be determined from the
partial differential equation
\begin{equation}
\label{diffEqG}
\nabla \times \nabla \times \bm{G}(\mathbf{r},\mathbf{r}',\omega)
-\frac{\omega^2}{c^2}\varepsilon (\mathbf{r},\omega)
\bm{G}(\mathbf{r},\mathbf{r}',\omega)=
\mathbf{\delta} (\mathbf{r}-\mathbf{r}')\bm{U},
\end{equation}
where $\bm{U}$ is the unit dyad. An important consequence of the
differential equation (\ref{diffEqG}) is the integral relation
\cite{SCHEEL/98}
\begin{equation}
\label{magicformula}
\int d^3\mathbf{s} \frac{\omega^2}{c^2}
\varepsilon_I(\mathbf{s},\omega) \bm{G}(\mathbf{r},\mathbf{s},\omega)
\bm{G}^+(\mathbf{r}',\mathbf{s},\omega) =
\mathrm{Im}\bm{G}(\mathbf{r},\mathbf{r}',\omega) .
\end{equation}

Quantization of this theory then
proceeds in the usual way \cite{PERINA}. First, a factor is split off
from the (classical) noise polarization,
\begin{equation}
\label{polarizNoise}
\mathbf{P}_N (\mathbf{r},\omega) = i \sqrt{\frac{\hbar
\varepsilon_0}{\pi} \varepsilon_I (\mathbf{r},\omega)}\,
\mathbf{f} (\mathbf{r},\omega ).
\end{equation}
One then identifies the dynamical variables
$\mathbf{f}(\mathbf{r},\omega)$ as the fundamental $\delta$ correlated
Gaussian random process and, upon quantization, replaces them by the
operator-valued bosonic vector field
$\hat{\mathbf{f}}(\mathbf{r},\omega)$ satisfying the equal-time
commutation relations
$\left[\hat{\mathbf{f}}(\mathbf{r},\omega),\hat{\mathbf{f}}^{\dagger}(\mathbf{r}',\omega')\right]$
$\!=$ $\!\delta(\mathbf{r}-\mathbf{r}')\delta(\omega-\omega')\bm{U}$.
The Hamiltonian of the system composed of electromagnetic field and
absorbing matter is
\begin{equation}
\label{FieldH}
\hat{H}_F = \int d^3 \mathbf{r} \int\limits_0^{\infty} d \omega \,
\hbar \omega \,\hat{\mathbf{f}}^{\dagger}(\mathbf{r},\omega)
\hat{\mathbf{f}}(\mathbf{r},\omega).
\end{equation}

The electromagnetic field operators can now be obtained in the
Schr\"odinger picture as
\begin{gather}
\label{eq:eschroedinger}
\hat{\mathbf{E}}(\mathbf{r}) =
\int\limits_0^{\infty} d \omega
\hat{\mathbf{E}}(\mathbf{r},\omega)+\mbox{H.c.},
\\
\label{ElectricFoper}
\hat{\mathbf{E}}(\mathbf{r},\omega) =
i \sqrt{\frac{\hbar}{\pi\varepsilon_0}} \frac{\omega^2}{c^2}
\int d^3 \mathbf{r}'\sqrt{\varepsilon_I (\mathbf{r}', \omega)}
\bm{G}(\mathbf{r},\mathbf{r}',\omega)
\hat{\mathbf{f}}(\mathbf{r}',\omega)
\end{gather}
and, using Faraday's law,
\begin{equation}
\label{magneticFoper}
\hat{\mathbf{B}}(\mathbf{r},\omega) = (i \omega)^{-1} \bm{\nabla} \times
\hat{\mathbf{E}}(\mathbf{r},\omega).
\end{equation}

An important feature of this theory is that it reproduces the correct
form of the fluctuation-dissipation theorem. Let the system of
electromagnetic field and absorbing matter be in thermal equilibrium at
some temperature $T$. Then the thermal correlation function of the
dynamical variables at temperature $T$ reads
\begin{equation}
\label{eq:correlator}
\langle \hat{\mathbf{f}}(\mathbf{r},\omega)
\hat{\mathbf{f}}^\dagger(\mathbf{r}',\omega') \rangle =
(\bar{n}_{\mathrm{th}}+1) \delta(\mathbf{r}-\mathbf{r}')
\delta(\omega-\omega') \bm{U},
\end{equation}
with the mean thermal photon number at frequency $\omega$
\begin{equation}
\bar{n}_{\mathrm{th}} = \frac{1}{e^{\hbar\omega/k_BT}-1}.
\end{equation}
From Eqs.~(\ref{ElectricFoper})---(\ref{eq:correlator}), together with
Eq.~(\ref{magicformula}), it follows that the thermal expectation value
of an anti-normally ordered product of magnetic field operators can be
written as
\begin{align}
\label{eq:fdt}
&
\langle \hat{\mathbf{B}}(\mathbf{r},\omega)
\hat{\mathbf{B}}^\dagger(\mathbf{r}',\omega') \rangle = \nonumber \\
&
\frac{\hbar\mu_0}{\pi}
\mathrm{Im}\left[ \overrightarrow{\bm{\nabla}} \times
\bm{G}(\mathbf{r},\mathbf{r}',\omega) \times
\overleftarrow{\bm{\nabla}} \right] (\bar{n}_{\mathrm{th}}+1)
\delta(\omega-\omega') .
\end{align}

Such a quantization model provides a valid description of
electromagnetic field in absorbing dielectric materials. In fact, it
has been shown in \cite{DUNG/98,SCHEEL/98} that the equal-time basic
commutation relations of QED are preserved. The electromagnetic
field is expressed in terms of the classical Green tensor
satisfying the Helmholtz equation (\ref{HelmholtzEq}), and the
continuum of the bosonic field variables
$\hat{\mathbf{f}}(\mathbf{r},\omega)$. All the information about the
dielectric matter is contained in the Green tensor via the
permittivity $\varepsilon (\mathbf{r},\omega)$.
For metals at low frequencies, the permittivity can be approximated by
the well-known Drude relation
\begin{equation}
\varepsilon(\omega) \approx \frac{2ic^2}{\omega^2\delta^2}
\end{equation}
with the skin depth $\delta$. Although such a relation is not strictly
consistent with causality as it has recently been pointed out
\cite{Tip}, it can be assumed to be valid in a restricted frequency
interval.

At this point it is necessary to point out the limitations of the
quantization scheme presented above. Note that the form of the
polarization, Eq.~(\ref{Polarization}), is valid only for strictly
locally responding materials. That is to say, we assume that the
elementary dipoles that give rise to the polarization are essentially
fixed in space. Certainly, for metals which can alternatively
described by a conductivity, this is not true as charge carriers can
move around freely for considerable distances. However, the locality
assumption can be upheld in situations in which the mean free path
length is much shorter than all the other length scales in the system
under consideration. While this is certainly true for ordinary metals
at room temperature and geometric length scales of several
micrometers, we do expect corrections due to spatially nonlocal
response (the anomalous skin effect) for metals or superconductors at
very low temperatures as considered in \cite{Scheel05,Henkel05}.

%%%%%%%%%%%%%%%%%%%%%%%%%%%%%%%%%%%%%%%%%%%%%%%%%%%%%%%%%%%%%%%%%%%%%%
\section{Spatial Decoherence}
\label{sec:dec}

Let us suppose we had an atom in one of two adjacent sites of an
optical lattice. The tunneling interaction allows the atom's wave
function to coherently spread over the neighboring site \cite{HAYCOCK}
where its state can be written in the occupation-number basis as
\begin{equation}
|\psi(t=0)\rangle_A = \frac{1}{\sqrt{2}}
\left( |1,0\rangle +|0,1\rangle \right).
\end{equation}
We take the time at which the equal superposition has been established
to be $t=0$ and assume for simplicity that no tunneling occurs at
later times, at least not at timescales shorter than the decoherence
time. This means that we imagine the tunneling interaction being
frozen over a certain time period. This assumption is justified
when considering proposals in which spatial atomic locations are used
to encode quantum information.

Atoms that are held close to microstructured surfaces experience
fluctuations of the electromagnetic field due to absorption in the
substrate material. In the case of a magnetic trap the atom is
subject to a constant magnetic field with strength $B_0$ in the
center of the trap. The magnetic sublevels are split due to the
Zeeman effect by the Larmor frequency $\omega_L=g_S\mu_B
B_0/\hbar$. A subset of these magnetic sublevels feel an
attractive potential towards regions of low magnetic field. In the
experiment reported in \cite{EXPATOMCHIP} $^{87}$Rb atoms are
initially pumped into the hyperfine state
$|F,m_F\rangle=|2,2\rangle$ in which they are trapped. However,
due to absorption in the surface  material and the resulting
quantum fluctuations, fluctuating magnetic fields cause the atoms
to evolve into states with lower magnetic quantum number $m_F$. In
sufficiently tight magnetic traps, also atoms in the
$|F,m_F\rangle=|2,1\rangle$ state are trapped.
Spin flips to even lower magnetic sublevels cause the atoms to be
expelled from the trap. In this case, spatial decoherence is no
more a matter of interest. Hence, it is sufficient to treat the
atomic system in a two-level approximation.

We focus on the Zeeman coupling of the atomic magnetic
moment to a fluctuating field represented by the Hamiltonian
\begin{equation}
\label{Hzeeman}
\hat H_Z = - \hat{\bm{\mu}} \cdot \hat{\textbf{B}}({\textbf{r}}_A),
\end{equation}
where the operator of the magnetic induction is given by
Eq.~(\ref{magneticFoper}), together with Eqs.~(\ref{eq:eschroedinger})
and (\ref{ElectricFoper}). The magnetic moment operator in
Eq. (\ref{Hzeeman}) associated with a transition $|i\ra\to|f\ra$ can
be written as $\hat{\bm{\mu}}=\bm{\mu} |i\ra \la f|+ \mbox{h.c.}$.
Since we assume the atom to be cooled into its electronic ground
state, there is no contribution of the angular momentum. Furthermore,
since the nuclear magnetic moment can be neglected because of the
ratio of the electron mass to the mass of the nucleus (see the
discussions in \cite{HENKEL/99,SPIN-FLIP}), the magnetic moment vector
is just proportional to the expectation value of the electronic spin
operator,
\begin{equation}
\bm{\mu} = g_S \mu_B \la i|\hat{\mathbf{S}}|f \ra ,
\end{equation}
where $\mu_B$ denotes the Bohr magneton, and $g_S\approx 2$ the
electron's $g$-factor. Inserting Eq.~(\ref{magneticFoper}) into
Eq.~(\ref{Hzeeman}), the Zeeman Hamiltonian can be written in the
rotating-wave approximation as \cite{SPIN-FLIP}
\begin{eqnarray}
\hat H_Z &= &
-\mu_B g_S \left[ \langle f|\hat{S}_q| i\rangle \hat{\xi}^\dagger
\hat{B}_q(\mathbf{r}_A) + \mbox{h.c.} \right]
\nonumber \\ &=&
-\mu_B g_S \left [\langle f|\hat{S}_q| i\rangle
\int\limits_0^\infty d \omega
\frac{\omega}{c^2}\sqrt{\frac{\hbar}{\varepsilon_0 \pi}}\,
\epsilon_{qpj}\, \partial_p \int d^3 \mathbf{s}
\right. \nonumber \\ &&
\left. \times
\sqrt{\varepsilon_I(\mathbf{s},\omega)}
G_{ji} ({\mathbf{r}}_A,{\mathbf{s}},\omega)
\hat{f}_i({\mathbf{s}},\omega)
\hat{\xi}^{+} +\mbox{h.c.} \right]
\nonumber \\
\label{Happrox}
\end{eqnarray}
where $\hat{\xi}=|f\rangle\langle i|$ denotes the atomic spin
lowering operator. Finally, the free atomic Hamiltonian can be written
in the two-level approximation used above as
\begin{equation}
\label{eq:Hatom}
\hat H_A = \hbar\omega_A \hat{\xi}_z =
\frac{1}{2} \hbar\omega_A (|i\rangle\langle i|-|f\rangle\langle f|),
\end{equation}
where the $\hat{\xi}$ obey the commutation rules
$[\hat{\xi}^{(\dagger)},\hat{\xi}_z]=\mp\hat{\xi}^{(\dagger)}$.

In order to analyze how this magnetic noise influences the
coherence of the state of our atom, we rewrite the initial atomic
state as
\begin{equation}
|\psi_A\rangle = \frac{1}{\sqrt{2}} \left( |i_1\rangle +
|i_2\rangle \right),
\end{equation}
where the labels $1, 2$ refer to the occupied site. Let us
consider a system composed of the two-level atom and a
fluctuating magnetic field initially in the vacuum state
$|0\rangle$, so that the total state of the atom-field system reads
\begin{eqnarray}
|\psi_{AF}\rangle =  \frac{1}{\sqrt{2}}\left (|i_1,0\rangle +
|i_2,0\rangle \right ).
\end{eqnarray}
The Hamiltonian describing the evolution of the combined
system is given by the sum of the three Hamiltonians
$\hat{H} = \hat{H}_F+ \hat{H}_A +\hat{H}_Z$, where
$\hat{H}_F$, and $\hat{H}_Z$, and $\hat{H}_A$ are given by
Eqs.~(\ref{FieldH}), (\ref{eq:Hatom}), and (\ref{Happrox}),
respectively. The system wave function at a certain time $t$ can be
written as \cite{DUNG/00}
\begin{eqnarray}
\lefteqn{
|\psi_{AF} (t)\rangle =
C_{i_1}(t) e^{-i \omega_A t/2} |i_1, 0 \rangle
+ C_{i_2}(t) e^{-i \omega_A t/2} |i_2, 0 \rangle }
\nonumber \\
&& +\int d^3\mathbf{r} \int\limits_0^\infty d\omega \,
C_{{f_1},m}({\mathbf{r}},\omega,t) e^{-i(\omega -\omega_A /2)t}
|f_1,1_m({\mathbf{r}},\omega)\rangle
\nonumber \\
&& +\int d^3\mathbf{r} \int\limits_0^\infty d\omega \,
C_{{f_2},m}({\mathbf{r}},\omega,t)
e^{-i(\omega -\omega_A /2)t} |f_2,1_m({\mathbf{r}},\omega)\rangle,
\nonumber \\
\label{SystemState}
\end{eqnarray}
where $|0\rangle$ and $|1_m({\mathbf{r}},\omega)\rangle$
denote the electromagnetic field vacuum and single-excitation states,
respectively. The Schr\"{o}dinger equation
$i\hbar\partial_t|\psi_{AF}(t)\rangle=\hat{H}|\psi_{AF}(t)\rangle$
yields ($a=1,2$)
\begin{eqnarray}
\label{CinizDot}
\lefteqn{
\dot{C}_{i_a}(t) =
\frac{i\mu_B g_S}{c^2\sqrt{\pi\hbar\varepsilon_0}}
\langle f|\hat{S}_q| i\rangle
\int d^3\mathbf{r}  \int\limits_0^\infty d\omega  }
\nonumber \\ && \times
\omega e^{-i(\omega-\omega_A)t}
\sqrt{\mathbf{\varepsilon}_I(\mathbf{r},\omega)}
\epsilon_{qpj} \partial_p G_{jm}(\mathbf{r}_a,\mathbf{r},\omega)
\nonumber \\ && \times
C_{f_a,m}(\mathbf{r},\omega,t) ,
\end{eqnarray}
\begin{eqnarray}
\label{eq:cfoft}
\lefteqn{
\dot{C}_{f_a,m}(\mathbf{r},\omega,t) =
\frac{i \mu_B g_S}{c^2 \sqrt{\pi \varepsilon_0 \hbar}}
\langle i|\hat{S}_q| f\rangle \omega e^{i(\omega -\omega_A)t} }
\nonumber \\ &&
\times \sqrt{\varepsilon_I(\mathbf{r},\omega)}
\epsilon_{qpj}\partial_p
G_{jm}^\ast(\mathbf{r}_a,\mathbf{r},\omega)
C_{i_a}(t).
\end{eqnarray}

We now substitute the result of formal integration of
$C_{f_a,m}(\mathbf{r},\omega,t)$ with the condition
$C_{f_a,m}(\mathbf{r},\omega,0)=0$ into
$\dot{C}_{i_a}(t)$, make use of the integral relation
(\ref{magicformula}), and obtain
\begin{equation}
\label{Ci1Dot}
\dot{C_{i_a}}(t)
=\int^t_0 dt' \,K_a(t-t')C_{{i_a}}(t'),
\end{equation}
where the integral kernel is
\begin{eqnarray}
\lefteqn{
K_a(t-t')=
-\frac{(\mu_B g_s)^2}{c^2 \pi \varepsilon_0 \hbar}\,
\langle f|\hat{S}_q| i\rangle\,
\langle i|\hat{S}_k| f\rangle }
\nonumber \\ && \times
\int\limits_0^\infty d\omega\, e^{-i(\omega -\omega_A)(t-t')}
\mathrm{Im}\left[ \overrightarrow{\bm{\nabla}}\times
\bm{G}(\mathbf{r}_a,\mathbf{r}_a,\omega) \times
\overleftarrow{\bm{\nabla}}\right]_{qk}.
\nonumber \\
\end{eqnarray}
We integrate both sides of Eq.~(\ref{Ci1Dot}) over $t$, and
change the order of integrations on the right-hand side we
derive
\begin{equation}
C_{i_a}(t)-C_{i_a}(0) = \int\limits_0^t dt' \,\bar{K}_a(t-t') C_{i_a}(t')
\end{equation}
with
\begin{eqnarray}
\lefteqn{
\bar{K}_a (t-t') =
\frac{(\mu_B g_S)^2}{c^2 \pi \varepsilon_0 \hbar}\,
\langle f|\hat{S}_q| i\rangle\,
\langle i|\hat{S}_k| f\rangle } \nonumber \\ && \hspace*{-3ex}\times
\int\limits_0^\infty d\omega\,
\frac{e^{-i(\omega -\omega_A)(t-t')} -1}{i(\omega -\omega_A)}
\mathrm{Im}\left[ \overrightarrow{\bm{\nabla}} \times
\bm{G}(\mathbf{r}_a,\mathbf{r}_a,\omega) \times
\overleftarrow{\bm{\nabla}} \right]_{qk} \nonumber \\
\end{eqnarray}
and the initial condition $C_{i_a}(0)=1$.
When the Markov approximation applies, i.e., when in  coarse
grained description of the atomic motion memory effects are
disregarded, we may let \cite{BARNETT}
\begin{equation}
\label{Markov}
\frac{\left (e^{-i(\omega -\omega_A)(t-t')} -1 \right )}
{i(\omega -\omega_A)}\rightarrow -\pi\delta(\omega-\omega_A) +i{\cal
P}\frac{1}{\omega-\omega_A}.
\end{equation}
Defining the coefficients
\begin{eqnarray}
\lefteqn{ \label{A1} \Gamma_a = 2 \left (\frac{(\mu_B g_S)^2 }{c^2
\varepsilon_0 \hbar} \right)\,\langle f|\hat{S}_q| i\rangle\,
\langle i|\hat{S}_k| f\rangle } \nonumber \\ && \times \mathrm{Im}
\left[ \overrightarrow{\bm{\nabla}}\times
\bm{G}(\mathbf{r}_a,\mathbf{r}_a,\omega_A) \times
\overleftarrow{\bm{\nabla}}\right]_{qk}
\end{eqnarray}
and
\begin{eqnarray}
\label{eq:deltaomega} \lefteqn{ \delta\omega_a = \left(
\frac{(\mu_B g_S)^2 }{c^2 \pi \varepsilon_0\hbar} \right) \langle
f|\hat{S}_q| i\rangle\, \langle i|\hat{S}_k| f\rangle } \nonumber
\\ && \times \mathcal{P} \int\limits^{\infty}_0 d\omega
\frac{\mathrm{Im}\left[\overrightarrow{\bm{\nabla}}\times
\bm{G}(\mathbf{r}_a,\mathbf{r}_a,\omega_A) \times
\overleftarrow{\bm{\nabla}} \right]_{qk}} {\omega - \omega_A},
\end{eqnarray}
we can write
$\bar{K}_a(t-t')=-\frac{1}{2}\Gamma_a+i\delta\omega_a$.
We finally obtain for the time evolution of the coefficients
$C_{i_a}(t)$
\begin{equation}
\label{Ci1Sol}
C_{i_a} (t) = \exp \left[ \left( -\frac{1}{2} \Gamma_a + i
\delta\omega_a \right) t \right] .
\end{equation}
The coefficients $\Gamma_a$ and $\delta\omega_a$ defined in
Eqs.~(\ref{A1}) and (\ref{eq:deltaomega}) represent the spin flip
rate and the line shift, respectively, and have been derived in a
similar fashion in \cite{SPIN-FLIP}. The spin flip lifetimes
$1/\Gamma_a$ have already been subject of major theoretical
\cite{Varpula,HENKEL/99,FOLMAN&Al,HENKEL/03,SPIN-FLIP,Henkel05,Scheel05}
and experimental \cite{EXPATOMCHIP,Vuletic,Cornell} investigations
which will not repeated here. In what follows, we will assume that
the line shift $\delta\omega_a$ caused by the interaction with the
quantized electromagnetic field is negligible. This can be seen as
follows. The Green function appearing in Eq.~(\ref{ElectricGreen}), as
well as the Fourier transform of the permittivity in
Eq~(\ref{Polarization}), plays the role of a response function and so
it satisfies the Kramers-Kronig relations for a complex-valued function
$g(\omega)=\mathrm{Re}[g(\omega)]+i \mathrm{Im} [g(\omega)]$
\cite{NUSSENZVEIG},
\begin{eqnarray}
\mathrm{Re}[g(\omega)]&=&\frac{1}{\pi}  \mathcal{P}
\int\limits^{\infty}_{-\infty} d\omega' \frac{
\mathrm{Im}[g(\omega)]}{\omega'-\omega},
\\
\mathrm{Im}[g(\omega)]&=&-\frac{1}{\pi} \mathcal{P}
\int\limits^{\infty}_{-\infty} d\omega' \frac{
\mathrm{Re}[g(\omega)]}{\omega'-\omega}.
\end{eqnarray}
The lower limit of the integral in Eq. (\ref{eq:deltaomega}) can be
extended to $-\infty$ with little error as the integrand is peaked
around $\omega_A$. Hence, Eq.~(\ref{eq:deltaomega}) can be rewritten as
\begin{eqnarray}
\label{eq:deltaomega2}
\lefteqn{
\delta\omega_a = \left(
\frac{(\mu_B g_S)^2 }{c^2 \varepsilon_0\hbar} \right) \langle
f|\hat{S}_q| i\rangle\, \langle i|\hat{S}_k| f\rangle } \nonumber
\\ && \times
\mathrm{Re}\left[\overrightarrow{\bm{\nabla}}\times
\bm{G}(\mathbf{r}_a,\mathbf{r}_a,\omega_A) \times
\overleftarrow{\bm{\nabla}} \right]_{qk} .
\end{eqnarray}
As we will see later, the line shift is of the same order of
magnitude as the spin flip rate. For typical experimental
realizations,
\cite{HENKEL/99,FOLMAN&Al,HENKEL/03,SPIN-FLIP,Henkel05,Scheel05,EXPATOMCHIP,Vuletic,Cornell},
this will be in the sub-Hz range. This means that $\delta
\omega_a$ can be neglected as it is extremely small when compared
to the spin flip transition frequency.

Now substituting Eq.~(\ref{Ci1Sol}) into the expression for
$\dot{C}_{f_a,m}(\mathbf{r},\omega,t)$, Eq.~(\ref{eq:cfoft}), we find
the formal solution
\begin{align}
& C_{f_a,m}(\mathbf{r},\omega,t)= \frac{i \mu_B g_S}{c^2 \sqrt{\pi
\varepsilon_0 \hbar}} \langle i|\hat{S}_q| f\rangle \omega
\sqrt{\varepsilon_I(\mathbf{r},\omega)} \epsilon_{qpj}\partial_p
\nonumber \\ & \times  G_{jm}^\ast(\mathbf{r}_a,\mathbf{r},\omega)
\int\limits_0^t dt' e^{i(\omega -\omega_A)t'} e^{-\frac{1}{2}
\Gamma_a t'}. \label{Cf1}
\end{align}
In order to find how the off-diagonal elements of the density
matrix decay, we trace the atomic density matrix over the field
and obtain
\begin{eqnarray}
\label{MatRhoA}
\varrho_A(t)  &=&
\langle 0|\varrho_{AF}(t)|0\rangle
\nonumber \\ &&
+ \sum_i \int d^3\mathbf{r} \int\limits_0^\infty d\omega \,
\langle 1_i(\mathbf{r},\omega)|\varrho_{AF}|
1_i(\mathbf{r},\omega)\rangle
\nonumber \\
&=&  \frac{1}{2} \left(
\begin{array}{cc}
\rho_{11}(t) & \rho_{12}(t) \\ \rho_{12}^\ast(t) & \rho_{22}(t) \\
\end{array}
\right),
\end{eqnarray}
where the matrix elements $\varrho_{ij}$ of the density matrix have to
be calculated from
\begin{eqnarray}
\label{rho1}
\varrho_{11}(t) &=& |C_{i_1}(t)|^2 + \sum_i\int d^3\mathbf{r}
\int\limits_0^\infty d\omega\,
|C_{f_1,m}({\mathbf{r}},\omega,t)|^2,
\nonumber \\ \\
\label{rho4}
\varrho_{22}(t) &=& |C_{i_2}(t)|^2 + \sum_i\int d^3\mathbf{r}
\int\limits_0^\infty d\omega \,
|C_{f_2,m}(\mathbf{r},\omega,t)|^2 ,
\nonumber \\ \\
\varrho_{12}(t) &=& C_{i_1}(t) C^\ast_{i_2}(t)
\nonumber \\
&& +\sum_i\int d^3\mathbf{r} \int\limits_0^\infty d\omega \,
C_{f_1,m}(\mathbf{r},\omega,t)\,C^\ast_{f_2,m}(\mathbf{r},\omega,t),
\label{rho2} \nonumber\\
\end{eqnarray}
First, it can be checked that the diagonal elements $\varrho_{11}(t)$ and
$\varrho_{22}(t)$ are properly normalized to
$\varrho_{11}(t)=\varrho_{22}(t)=1$
by inserting Eqs.~(\ref{Ci1Sol}) and (\ref{Cf1}) together with
Eq.~(\ref{magicformula}) into Eqs.~(\ref{rho1}) and (\ref{rho4}),
respectively. Thus, as a consistency check we find that
$\mathrm{Tr}[\varrho_A]=1$.
We can then calculate the off-diagonal elements of the
density matrix as
\begin{eqnarray}
\label{eq:offdiagonal}
\lefteqn{
\varrho_{12}(t) = e^{-\Gamma_{12} t}
+2\left( 1-e^{-\Gamma_{12} t} \right)
\frac{ \left( \mu_B g_S \right)^2}{c^2 \varepsilon_0 \hbar} }
\nonumber \\ && \times
\langle i|\hat{S}_q| f\rangle\langle f|\hat{S}_k| i\rangle
\frac{\mathrm{Im} \left[ \overrightarrow{\bm{\nabla}} \times
\bm{G}(\mathbf{r}_2,\mathbf{r}_1,\omega_A) \times
\overleftarrow{\bm{\nabla}} \right]_{kq}}
{\Gamma_{12}} \nonumber \\
\end{eqnarray}
where $\Gamma_{12}=(\Gamma_1+\Gamma_2)/2$ is the arithmetic mean of
the spin flip rates, Eq.~(\ref{A1}), at both sites.
Note that the Hermiticity of the density matrix $\varrho_A(t)$
follows from the reciprocity theorem applied to the dyadic Green
function which yields $\bm{G}(\mathbf{r}_1,\mathbf{r}_2,\omega_A)$
$\!=$ $\!\bm{G}^T(\mathbf{r}_2,\mathbf{r}_1,\omega_A)$.

Equation~(\ref{eq:offdiagonal}) constitutes the main result of our
paper. It provides, via the Green function
$\bm{G}(\mathbf{r}_2,\mathbf{r}_1,\omega_A)$, an elegant way to assess
the loss of spatial coherence for arbitrarily shaped
substrates. Recalling the expression for the fluctuation-dissipation
theorem,  Eq.~(\ref{eq:fdt}), it follows that
Eq.~(\ref{eq:offdiagonal}) can be rewritten as
\begin{eqnarray}
\label{eq:coherence}
\lefteqn{
\varrho_{12}(t) = e^{-\Gamma_{12}t} +\left( 1-e^{-\Gamma_{12}t}
\right) } \nonumber \\ && \hspace*{-3ex} \times
\frac{\langle i|\hat{S}_q| f\rangle\langle f|\hat{S}_k| i\rangle
\int_0^\infty d\omega \la
\hat{B}_k(\mathbf{r}_2,\omega_A)\hat{B}_q^\dagger(\mathbf{r}_1,\omega)\ra}%
{\langle i|\hat{S}_q| f\rangle\langle f|\hat{S}_k| i\rangle
\int_0^\infty d\omega \la
\hat{B}_k(\mathbf{r}_1,\omega_A)\hat{B}_q^\dagger(\mathbf{r}_1,\omega)\ra}
\nonumber \\ && \hspace*{-3ex} \equiv e^{-\Gamma_{12}t} +\left(
1-e^{-\Gamma_{12}t} \right) S \left (
\mathbf{r}_1,\mathbf{r}_2,\omega_A \right )
\end{eqnarray}
in terms of the magnetic cross-correlation tensor
$\la\hat{\mathbf{B}}(\mathbf{r},\omega)\hat{\mathbf{B}}^\dagger(\mathbf{r}',\omega')\ra$.
This means that the imaginary part of the (magnetic) Green
function is proportional to the spatial coherence function of the
fluctuating magnetic field
\cite{CARMINATI/99,HENKEL_CARMINATI/00,Henkel01}.

Note that, although the calculations have been performed for surfaces
held at zero temperature, the extension to finite temperatures is
trivial. Indeed, it is seen from Eq.~(\ref{eq:fdt}) that the spatial
coherence functions as well as the spin-flip rates simply have to be
multiplied by the factor $(\bar{n}_{\text{th}}+1)$ to account for
thermal fluctuations.

Equation~(\ref{eq:offdiagonal}), or equivalently,
Eq.~(\ref{eq:coherence}), consists of two parts. The first is a
(spatially local) exponential decay that describes the effect of
the transition from the initial spin state $|i\rangle$ to the
final spin state $|f\rangle$. The second term is a (spatially
nonlocal) non-exponential term which is proportional to the
spatial coherence function. It should be noted that, in a model in
which more than a two-level transition is considered, after this
time a transition to even lower-lying hyperfine spin states are
likely. However, in our two-level approximation these flips are
not taken into consideration.

%%%%%%%%%%%%%%%%%%%%%%%%%%%%%%%%%%%%%%%%%%%%%%%%%%%%%%%%%%%%%%%%%%%%%%
\section{Planar multilayer substrates}
\label{sec:planar}

Up until now, the derivation of all formulas were valid for
arbitrary substrate geometries. A particular geometric arrangement
is fixed by defining the correct boundary conditions for the
dyadic Green function $\bm{G}(\mathbf{r},\mathbf{s},\omega)$. In
this section, we will concentrate on the simplest but
experimentally important realization in terms of planar multilayer
dielectrics. In what follows, we will focus on the spatially nonlocal
term in Eq.~(\ref{eq:offdiagonal}) only. In particular, we
notice that this is equivalent to taking the long-time limit of
Eq.~(\ref{eq:offdiagonal}). Hence, for now we consider only
\begin{eqnarray}
\label{eq:ratio} \lefteqn{ S \left (
\mathbf{r}_1,\mathbf{r}_2,\omega_A \right )= 2 \frac{ \left( \mu_B
g_S\right)^2}{c^2 \varepsilon_0\hbar} \langle i|\hat{S}_q|
f\rangle\langle f|\hat{S}_k| i\rangle } \nonumber \\ && \times
\frac{\mathrm{Im}\left[ \overrightarrow{\bm{\nabla}} \times
\bm{G}(\mathbf{r}_2,\mathbf{r}_1,\omega_A) \times
\overleftarrow{\bm{\nabla}} \right]_{kq}}{\Gamma_{12}},
\end{eqnarray}
which had previously been derived in connection with spatial
decoherence of matter waves in \cite{Henkel01}. Note that in a
planar geometry in which the atom is held at a fixed distance to
the material surface, the spin flip rates $\Gamma_i$ coincide due
to translational invariance, i.e.
$\Gamma_{12}\equiv\Gamma_1=\Gamma_2$. Note also that
Eq.~(\ref{eq:ratio}) is temperature-independent.

Let us first consider a half-space filled with a dielectric or
metal of dielectric permittivity $\varepsilon(\omega)$ (see the
discussion in Sec.~\ref{sec:QED}). We evaluate the spin matrix
elements for the transition from one hyperfine ground state to another
by the basis states through the Clebsch--Gordon coefficients
$|F,m_F\rangle=\sum_{m_Sm_I}C_{Fm_F}^{m_Sm_I} |m_S,m_I\rangle$. For
the $^{87}$Rb ground state transition $|2,2\rangle \to|2,1\rangle$,
the non-zero matrix elements are $|\langle i|\hat{S}_{y,z}|f\rangle|=1/4$.
The dyadic Green function for such a situation can
be found in \cite{LiLW94,Chew,TOMAS,DUNG/98}. We have collected
some of the formulas in Appendix~\ref{sec:green}. Note that in the
expressions for the components of the generalized reflection
coefficient, Eq.~(\ref{eq:coefficients}), the common factor
$e^{ik_{1z}(z+z')}\equiv e^{2ik_{1z}d}$ can be approximated by
$e^{-2d|k_\||}$ because the transition wavelength,
$\lambda=c/(2\pi\omega)$, is the by far biggest length scale in
the system such that the approximation $k^2_{1z}\approx-k_\|^2$
holds. Then, by going over to polar co-ordinates in the
two-dimensional Fourier transform in Eq.~(\ref{eq:Weyl}),
$\mathbf{k}_\|=(k_x,k_y)\mapsto(K\cos\varphi,K\sin\varphi)$ and
$d^2\mathbf{k}_\|\mapsto KdK\,d\varphi$. We can thus write
$\Gamma_{12}$ after integration over $\varphi$ as
\begin{equation}
\label{Gamma12} \Gamma_{12} = \frac{(\mu_B
g_S)^2}{8c^2\varepsilon_0 \hbar}\, 3 \pi \int
\frac{K^2dK}{(2\pi)^2} \frac{e^{-2Kd}}{2}
\mathrm{Im}[r^{TE}_{12}].
\end{equation}
It is worth noting at this point that the line shift
$\delta\omega_a$ in Eq.~(\ref{eq:deltaomega2}) can be computed as
in Eq.~(\ref{Gamma12}) by replacing $\mathrm{Im}[r^{TE}_{12}]$
with $\mathrm{Re}[r^{TE}_{12}]$. Moreover, it is easily seen that
both $\Gamma_a$ and $\delta \omega_a$ are of the same order.

Let us assume that an atom is located at a distance $d$ away
from the planar interface which we describe by its skin depth
$\delta$. In our example, we have chosen an aluminium substrate with
$\delta=110\mu$m and an atomic transition frequency as
$f=560$kHz. Furthermore, the atom can be in two distinct
positions with a lateral separation $l$.
\begin{figure}[ht]
\begin{center}
\includegraphics[width=8.2cm]{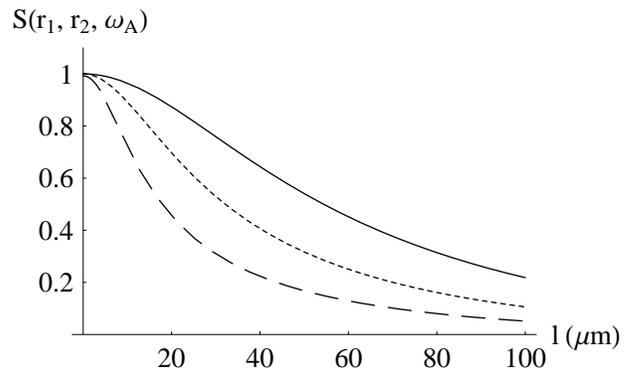}
\end{center}
\caption{\label{fig:Ratio} Spatial coherence function of the
fluctuating magnetic field $S \left (
\mathbf{r}_1,\mathbf{r}_2,\omega_A \right )$,
Eq.~(\ref{eq:ratio}), as a function of the lateral separation $l$
in $\mu$m with the parameters $f=560$~kHz, $\delta=110\,\mu$m for
three different distances from the surface: $d=20\,\mu$m (solid
line), $d=10\,\mu$m (dotted line), and $d=5\,\mu$m (dashed line).}
\end{figure}
In Fig.~\ref{fig:Ratio} we show the decay of the spatial coherence
as measured by the function $S \left (
\mathbf{r}_1,\mathbf{r}_2,\omega_A \right )$ for varying
separation $l$ in $\mu$m for three different atom-surface
distances $d$. As a function of separation, the decay of the
spatial coherence starts off rather slowly. We attribute this
behaviour to the fact that for separations below the coherence
length of the magnetic-field fluctuations the spin flip is driven
coherently at both sites.

In order to investigate the small-separation limit in some more
detail, we take a closer look at the Weyl expansion of the scattering
Green tensor $\bm{R}^{(12)}(\mathbf{r},\mathbf{r}',\omega)$,
Eq.~(\ref{eq:Weyl}), which is the by far dominant contribution
compared with the free-space Green function. The separation $l$ is
nothing but $l=|\bm{\varrho}-\bm{\varrho}'|$ and serves as a parameter
in the integral. Hence, we can expand the exponential
$e^{i\mathbf{k}_\|\cdot(\bm{\varrho}-\bm{\varrho}')}$ in
Eq.~(\ref{eq:Weyl}) into powers of $l$ and evaluate each term
seperately. The zeroth-order coefficient trivially leads to the spin
flip rate $\Gamma_{12}$. The contribution from terms proportional to $l$
vanish identically due to the symmetry of the generalized reflection
coefficients $R_{ij}^{(12)}$ with respect to the wave-vector
components $\mathbf{k}_\|$ in the $(x,y)$-plane.
In fact, all odd powers of $l$ vanish because of that symmetry.

Hence, the lowest non-vanishing power is $l^2$. It is
straightforward to find analytical expressions for the
spatial coherence in that limit by converting the additional
factor $K^2$ coming from the expansion of the exponential in
Eq.~(\ref{eq:Weyl}) into a parameter differentiation with respect
to the atom-surface distance $d$. That is, we make the replacement
$K^2\mapsto\frac{1}{4}\frac{\partial^2}{\partial d^2}$.
In this way we find that
\begin{equation}
\label{eq:ratio2} S \left ( \mathbf{r}_1,\mathbf{r}_2,\omega_A
\right ) = \frac{1}{\Gamma_{12}} \left( \Gamma_{12}-\frac{5
l^2}{96} \frac{\partial^2}{\partial d^2}\Gamma_{12} \right) +{\cal
O}(l^4).
\end{equation}
In certain asymptotic regimes in which $\Gamma_{12}$ can be expressed
as a monomial $\propto d^{-n}$ of the atom-surface distance $d$ (see,
e.g. \cite{HENKEL/99,Henkel01,Scheel05}), Eq.~(\ref{eq:ratio2}) can be
rewritten in the form
\begin{equation}
S \left ( \mathbf{r}_1,\mathbf{r}_2,\omega_A \right
)=1-\frac{5n(n+1)l^2}{96d^2} +{\cal O}(l^4).
\end{equation}
In addition to the planar half-space we consider the experimentally
relevant situation in which a thin metallic layer of thickness $h$ has
been brought onto a dielectric substrate. The generalized Fresnel
coefficient for this three-layer system is given in Eq.~(\ref{eq:Fresnel3}).
In the limit of thick films ($\delta,h\gg d$) the asymptotic behaviour
of the spin flip rate is $\Gamma_{12}\propto 1/d$
\cite{HENKEL/99,Scheel05} whereas for thin films ($\delta\gg d\gg h$)
we have $\Gamma_{12}\propto 1/d^2$ \cite{Henkel01,Scheel05}. Thus, we
finally obtain the small-$l$ limit of Eq.~(\ref{eq:offdiagonal}) as
\begin{equation}
\label{eq:smallsep} \varrho_{12}(t)=1-\frac{5\alpha l^2}{48d^2}
\left( 1-e^{-\Gamma_{12}t} \right) +{\cal O}(l^4)\,,
\end{equation}
where $\alpha=1$ for thick films and $\alpha=3$ for thin films. It
is interesting to note that the fall-off is three times faster for
thin films than for thick films which we attribute to the fact
that in thick films it is more likely to drive spin-flips
coherently.

In order to see how the time scale is related to the expected lifetime
we can expand the exponential in Eq.~(\ref{eq:smallsep}) for short
times as
\begin{equation}
\label{eq:smalltime} | \varrho_{12}(t)-\varrho_{12}(0)|\cong
\frac{5\alpha l^2}{48d^2} \left ( \frac{t}{\tau}\right )+ {\cal
O}(t^2)
\end{equation}
where $\varrho_{12}(0)=1$ and $\tau = \Gamma^{-1}_{12}$. The
left-hand side in Eq.~(\ref{eq:smalltime}) can be thought as a
proper measure of decoherence due to spin flips in terms of
physical parameters such as the spin-flip lifetime $\tau$, the
separation $l$ and the distance from the surface $d$. This means
that it is possible to maximize those experimental parameters
while the decoherence rate is under control. Hence,
Eq.~(\ref{eq:smalltime}) turns out to be particularly interesting
from the quantum information point of view when a certain degree
of spatial coherence has to be maintained.

\begin{figure}[ht]
\centerline{\includegraphics[width=8cm]{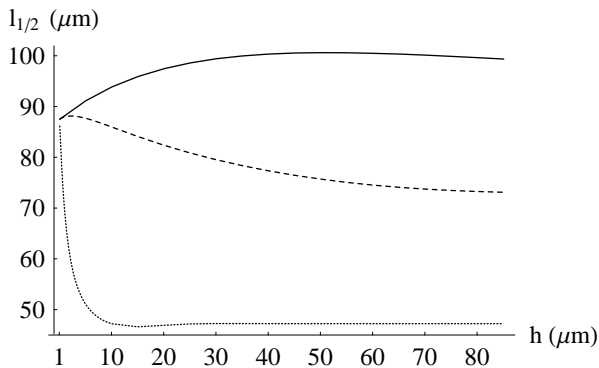}}
\caption{\label{fig:halflife} Lateral separation $l_{1/2}$ after
which spatial coherence has dropped to half its initial value as a
function of the layer thickness $h$. The skin depth was varied
from $\delta=100\mu$m (solid line) to $\delta=50\mu$m (dashed
line) and $\delta=10\mu$m (dotted line). The atom-surface distance
was $d=50\mu$m and all other parameters as in
Fig.~\ref{fig:Ratio}.}
\end{figure}

For larger separations, however, it is difficult to find analytical
approximations and one has to resort to numerical evaluations of
the Fourier transform (\ref{eq:Weyl}). It is interesting to see at
which separation $l_{1/2}$, as a function of the other length
parameters in the system, the spatial coherence drops to half its
initial value which could be taken as a measure of robustness.
In Fig.~\ref{fig:halflife} we show the dependence of $l_{1/2}$ on the
thickness $h$ of the intermediate layer.
In our calculations, we assumed a transition frequency of
$f=560$kHz. We have plotted $l_{1/2}$ for
three different skin depths: $\delta=100\mu$m (solid line,
corresponding to a good conductor such as Al of Cu at room
temperature), $\delta=50\mu$m (dashed line), and $\delta=10\mu$m
(dotted line). Although the latter two skin depth values are not
realistic for materials at room temperature, at kryogenic temperatures
these values can be achieved. For example, just above its critical
temperature of $T_c=9.2$K, pure niobium shows a skin depth of only
$\delta=15\mu$m at $f\lesssim 1$MHz \cite{Casalbuoni}.

In Fig.~\ref{fig:halflife} it is clearly seen that for skin depths
smaller than the atom-surface distance (dotted line), the robustness
of spatial coherence drops dramatically with increasing substrate
thickness $h$ until $h\sim\delta$. This can be understood when noting
that by increasing the thickness of the intermediate layer one
increases the number of fluctuating dipoles that can cause the spin
flip. Any further increase beyond $h\sim\delta$ does not change much
because fluctuations would not reach the substrate surface. Note also
that the coherence length $l_{1/2}$ levels out roughly at the value of
the skin depth, $l_{1/2}\sim\delta$.

For skin depths equal (dashed line in Fig.~\ref{fig:halflife}) or
larger than the atom-surface distance (solid line) spatial coherence
is robust over a wide range of substrate thicknesses $h$. Only for
$h\gtrsim\delta$ does the coherence length decrease towards the
atom-surface distance.

%%%%%%%%%%%%%%%%%%%%%%%%%%%%%%%%%%%%%%%%%%%%%%%%%%%%%%%%%%%%%%%%%%%%%%
\section{Conclusions}
\label{sec:conclusions}

In summary, we have investigated loss of spatial coherence of
atomic superpositions due to thermally driven spin flips. The
consistent quantization of the electromagnetic field in absorbing
dielectrics and metals allowed us to employ a first-principles
approach to decoherence in this particularly simple physical
system. The quantization scheme is based on the source-quantity
representation of the electromagnetic field in terms of the dyadic
Green function of the associated classical scattering problem and
a bosonic vector field that serves as the dynamical variables of
the theory. The Green function contains, via the dielectric
permittivity, all information about the geometric arrangement and
material properties of the substrate. Because the theory, starting
already with Eq.~(\ref{Polarization}), is strictly valid only for
spatially \textit{locally} responding materials, we stress again
that spatially nonlocal effects --- which could be non-negligible
for small skin depths (i.e. large conductivities) and small
atom-surface distances --- have not been considered.

The interaction dynamics between atomic spin and electromagnetic
field has been described in the Schr\"odinger picture and the
Markov approximation which led to the result for the time
evolution of the off-diagonal matrix element (or coherence)
$\varrho_{12}(t)$ of the single-particle density matrix,
Eq.~(\ref{eq:offdiagonal}). The spatially nonlocal part,
Eq.~(\ref{eq:ratio}), agrees with previously obtained results
\cite{Henkel01} for spatial decoherence of matter waves. It should
be noted that both Eqs.~(\ref{eq:offdiagonal}) and
(\ref{eq:ratio}) are valid for arbitrary geometrical arrangements
of substrate materials.

For planarly multilayered substrates the dyadic Green function is
explicitly known \cite{TOMAS,DUNG/98,Chew,LiLW94} and the main
formulas presented in Appendix~\ref{sec:green}. For small lateral
separation $l$ of the atom's two possible positions we found that
the spatial coherence decreases quadratically with $l$ and
inversely proportional to the squared atom-surface distance $d$
[Eq.~(\ref{eq:smallsep})]. For larger separations, a numerical
study of a three-layer system showed that the coherence length
$l_{1/2}$, defined to be the separation after which the coherence
decays to half its initial value, converges for thick intermediate
layers to roughly the atom-surface distance $d$.

We believe that these results are important for the design of
microstructured devices in which spatial coherences are used to
encode quantum information. In particular Eq.~(\ref{eq:smalltime})
shows how the decoherence rate depends on experimental parameters
such as lifetime, lateral separation and atom-surface distance.
They can be tuned in order to fall within a given tolerance rate
for the degree of decoherence. Therefore, the theoretical results
presented here may be useful in the physical realization of atomic
traps where a certain degree of spatial coherence has to be
maintained in order to be able to perform some kind of error
correction.

%%%%%%%%%%%%%%%%%%%%%%%%%%%%%%%%%%%%%%%%%%%%%%%%%%%%%%%%%%%%%%%%%%%%%%
\acknowledgments
This work was financially supported by the UK Engineering and Physical
Sciences Research Council (EPSRC) and the CONQUEST programme of the
European commission.
%%%%%%%%%%%%%%%%%%%%%%%%%%%%%%%%%%%%%%%%%%%%%%%%%%%%%%%%%%%%%%%%%%%%%%
\appendix
\section{Green function for planar multilayers}
\label{sec:green}

We briefly review the calculation of the Green function of planar
multilayers as it can be found in
\cite{TOMAS,DUNG/98,Chew,LiLW94}. The dyadic Green function for the
electric field scattering off a material interface can always be
decomposed into
\begin{equation}
\bm{G}(\textbf{r},\textbf{r}',\omega) = \left\{
\begin{array}{l}
\bm{G}^{(1)}(\textbf{r},\textbf{r}',\omega)
+ \bm{R}^{(12)}(\textbf{r},\textbf{r}',\omega)\,;
\textbf{r},\textbf{r}'\in {\cal V}_1 \\
\bm{T}^{(12)}(\textbf{r},\textbf{r}',\omega)\,;
\textbf{r}\in{\cal V}_1\,,\textbf{r}'\in {\cal V}_2
\end{array}
\right.
\end{equation}
where $\bm{G}^{(1)}(\textbf{r},\textbf{r}',\omega)$ denotes the
solution to the inhomogeneous Helmholtz equation with the source in
region ${\cal V}_1$ which in our case is vacuum with
$\varepsilon_1(\omega)\equiv 1$. The two (double-sided transverse)
scattering parts $\bm{R}^{(12)}(\textbf{r},\textbf{r}',\omega)$ and
$\bm{T}^{(12)}(\textbf{r},\textbf{r}',\omega)$ have to be introduced to
satisfy the boundary conditions for the electromagnetic fields at the
interface and describe the reflection and transmission parts of the
total scattering Green function, respectively. These scattering Green
functions satisfy the homogeneous Helmholtz equation. In our case, we
only need to concentrate on the reflection part
$\bm{R}^{(12)}(\textbf{r},\textbf{r}',\omega)$.

The translational invariance in two spatial directions, say in the
$(x,y)$-plane, allows one to write the Green function in terms of its
Weyl expansion
\begin{equation}
\label{eq:Weyl}
\bm{R}^{(12)}(\textbf{r},\textbf{r}',\omega) = \int
\frac{d^2\textbf{k}_\|}{(2\pi)^2}
\bm{R}^{(12)}(\textbf{k}_\|,\omega;z,z')
e^{i\textbf{k}_\|\cdot(\bm{\rho}-\bm{\rho}')}
\end{equation}
[$\bm{\rho}=(x,y)$] where $\textbf{k}_\|=(k_x,k_y)$ is the wave-vector
in the $(x,y)$-plane. The matrix components of
$\bm{R}^{(12)}(\textbf{k}_\|,\omega;z,z')$ can be read off from
\cite{DUNG/98} as (here we omit the arguments to enhance readability)
\begin{eqnarray}
\label{eq:coefficients}
R_{xx}^{(12)} &=& \frac{i}{2k_{1z}} e^{ik_{1z}(z+z')}
\left[ -r^{TM}_{12} \frac{k_{1z}^2 k_x^2}{k_1^2 k_\|^2}
+r^{TE}_{12} \frac{k_y^2}{k_\|^2} \right] \,,\nonumber \\
R_{xy}^{(12)} &=& \frac{i}{2k_{1z}} e^{ik_{1z}(z+z')}
\left[ -r^{TM}_{12} \frac{k_{1z}^2 k_xk_y}{k_1^2 k_\|^2}
-r^{TE}_{12} \frac{k_xk_y}{k_\|^2} \right] \,,\nonumber \\
R_{xz}^{(12)} &=& \frac{i}{2k_{1z}} e^{ik_{1z}(z+z')}
\left[ r^{TM}_{12} \frac{k_{1z} k_x}{k_1^2} \right] \,,\nonumber \\
R_{zz}^{(12)} &=& \frac{i}{2k_{1z}} e^{ik_{1z}(z+z')}
\left[ r^{TM}_{12} \frac{k_\|^2}{k_1^2} \right] \,,
\end{eqnarray}
where $k_i^2=\frac{\omega^2}{c^2}\varepsilon_i(\omega)$ and
$k_{iz}^2=k_i^2-k_\|^2$. The remaining matrix elements can be deduced
from Eq.~(\ref{eq:coefficients}) by replacement rules such as
$R_{yy}^{(12)}=R_{xx}^{(12)}(k_x \leftrightarrow k_y)$ and the
reciprocity condition
$\bm{R}^{(12)}(\textbf{r},\textbf{r}',\omega)$ $\!=$
$\!\bm{R}^{(12)T}(\textbf{r}',\textbf{r},\omega)$ which yields
$\bm{R}^{(12)}(\textbf{k}_\|,\omega;z,z')$ $\!=$
$\!\bm{R}^{(12)T}(-\textbf{k}_\|,\omega;z',z)$.

The functions $r^{TE}_{12}$ and $r^{TM}_{12}$ denote the usual Fresnel
reflection coefficients for TE and TM waves, respectively, and are
defined by
\begin{equation}
\label{eq:fresnel}
r^{TE}_{12} = \frac{k_{1z}-k_{2z}}{k_{1z}+k_{2z}} \,,\quad
r^{TM}_{12} = \frac{\varepsilon_2(\omega) k_{1z}
- \varepsilon_1(\omega)k_{2z}}{\varepsilon_2(\omega) k_{1z}-
\varepsilon_1(\omega)k_{2z}} \,.
\end{equation}
The Fresnel coefficients obey certain recursion relations that permit
one to calculate the dyadic Green function for arbitrarily
multi-layered materials \cite{Chew,TOMAS,LiLW94}. In particular, the
generalized Fresnel coefficient for a three-layer geometry reads (for
both TE and TM polarizations)
\begin{equation}
\label{eq:Fresnel3}
\tilde{r}_{12} = \frac{r_{12}+r_{23}e^{2ik_{2z}h}}%
{1-r_{21}r_{23}e^{2ik_{2z}h}}
\end{equation}
where $h$ is the thickness of the intermediate layer $2$. This
relation has been used in the numerical calculations throughout the paper.

%%%%%%%%%%%%%%%%%%%%%%%%%%%%%%%%%%%%%%%%%%%%%%%%%%%%%%%%%%%%%%%%%%%%%%

\end{document}